# Structural and magnetic properties of ZnO:TM (TM: Co,Mn) nanopowders

D. Rubi[a]*, A. Calleja[b], J. Arbiol[c], X.G. Capdevila[d], M. Segarra[d], Ll. Aragonès[b] and J. Fontcuberta[a]

[a] *Institut de Ciència de Materials de Barcelona (CSIC), Campus UAB, E08193, Belaterra, Spain*
[b] *Quality Chemicals, Fornal 35, Pol. Ind. Can Comelles Sud, Esparreguera, 08292, Spain*
[c] *TEM-MAT, Serveis Cientificotècnics, Universitat de Barcelona, Solé i Sabraís 1-3, Barcelona 08028, Spain*
[d] *Facultat de Química, Universitat de Barcelona, Martí i Franquès 1, 08028, Barcelona, Spain*

We report on the structural and magnetic characterization of $Co_{0.1}Zn_{0.9}O$ and $Mn_{0.1}Zn_{0.9}O$ nanopowders obtained by a soft chemistry route. We show that those samples fired at low temperatures display a ferromagnetic interaction that can not be attributed to the presence of impurities. A magnetic aging mechanism is observed, reflecting the key role played by defects in the stabilization of ferromagnetism in this kind of diluted magnetic semiconductors.

* Corresponding author. Tel.: +34 935 801 853; fax: +34 935 805 729.
  *E-mail address*: drubi@icmab.es

## 1. Introduction

Since early theoretical claims [1] proposing room-temperature ferromagnetism (FM) for TM:ZnO (where TM is a transition metal such as Co or Mn) diluted magnetic semiconductors, lot of efforts have been devoted to the study of these compounds, obtaining, quite often, contradictory results. The literature shows reports claiming for intrinsic ferromagnetism [2], paramagnetism [3], or extrinsic ferromagnetism (arising, for example, from the segregation of $Co^0$ clusters [4] or $(ZnMn)_2O_{3-\delta}$ [5]). Recently, the role of defects (namely, shallow donors or acceptors) in the stabilization of the ferromagnetic interaction has been convincingly put forward by Gamelin et al. [6], showing the possibility of controlling, by means of different chemical perturbations, the n-type and p-type defects necessary to mediate the ferromagnetic coupling of ZnO:Co and ZnO:Mn, respectively.

Here we report on the structural and magnetic properties of $Co_{0.1}Zn_{0.9}O$ and $Mn_{0.1}Zn_{0.9}O$ nanopowders, obtained by a soft chemistry route and further calcinated at different temperatures (i.e.: 300ºC-600ºC). It will be shown that those samples treated at low temperatures display a ferromagnetic behaviour which does not appear to be related to the presence of impurities, while those treated at higher temperatures are fully paramagnetic.

## 2. Experimental

$Co_{0.1}Zn_{0.9}O$ and $Mn_{0.1}Zn_{0.9}O$ powders were obtained by means of the acrilamide polymerization method, as described in detail elsewhere [7]. The xerogels obtained after self-combustion are fired in air in a muffle furnace at temperatures between 300ºC and 600ºC. Co-doped and Mn-doped powders displayed green and brown colours, respectively, both turning darker when increasing the firing temperature. We will show results corresponding to samples of $Co_{0.1}Zn_{0.9}O$ and $Mn_{0.1}Zn_{0.9}O$ fired at 300ºC (labelled as ZC1_a and ZM1_a, respectively) and 600ºC (labelled as ZC1_d and ZM1_d). Structural characterization was performed by X-ray diffraction (Siemens D-5000, $CuK_{\alpha 1, \alpha 2}$ radiation), and transmission electron microscopy (FEG TEM Jeol 2010F). The cationic composition of the samples was checked by means of energy dispersive X-ray spectroscopy (EDS), finding in all cases Mn and Co contents in agreement with the nominal composition, within the error of the technique. Magnetic measurements were performed by using a standard Quantum Design Superconductor Quantum Interference Device (SQUID) magnetometer.

3. Results

Figure 1 shows the high resolution XRD patterns corresponding to both Co- and Mn doped samples, showing the presence of the characteristic reflections of the wurtzite-like hexagonal structure of ZnO. For those samples treated at 600ºC, it is evident the presence of additional peaks (labeled as * and o) that reflect the segregation of secondary phases. In the case of Co-doped samples, this phase can be identified as a $Zn_xCo_{3-x}O_4$ spinel, while for Mn-doped samples the spurious peaks nicely match with those of $ZnMoO_3$. In both cases, it is evident that increasing the calcination temperature favors the phase segregation. The calculated cell parameters of the wurtzite-like phase of sample ZM1_a resulted a=3.262Å and c=5.219 Å, while those corresponding to sample ZM1_d are a=3.253Å and c=5.214Å, respectively. Bearing in mind the $Mn^{2+}$ and $Zn^{2+}$ ionic radii (0.66 Å and 0.60Å, respectively), and the cell parameters of pure

ZnO (a=3.250Å y c=5.207 Å) it is clear that for lower firing temperatures Mn substitutes $Zn^{2+}$ in the ZnO matrix, inducing a cell expansion as a consequence of its larger ionic size. Additionally, the observed shrink of the cell parameters when increasing the calcination temperature reflects the migration of Mn cations from the wurzite structure to the $ZnMnO_3$ impurity. In the case of Co-doped samples the estimated cell parameters of ZC1_a and ZC1_d were in both cases a=3.250Å and c=5.203Å, which are very close to the ZnO characteristic values. This should be related to the similarity between $Co^{2+}$ and $Zn^{2+}$ ionic radii (0.58 Å and 0.60 Å, respectively), which leads into an almost unmodified unit cell upon Co substitution.

The valence states of both Co and Mn ions in ZC1_a and ZM1_a samples were checked by means of optical absorption and X-ray photoemission (to be reported elsewhere). Data indicates the presence of $Co^{2+}$ and $Mn^{2+}$ cations in tetrahedral coordination. This reinforces the idea that both Co and Mn replace Zn ions in the wurtzite structure.

Figure 2(a) displays a TEM image corresponding to sample ZM1_a. It can be observed the presence of a porous network of nanometric particles, with sizes of about ~40-50nm. Figure 2(b) shows the same sample in an amplified scale. Interestingly enough, it can also be appreciated the existence of smaller rounded particles with sizes in the range ~5-10nm, which, based on detailed high-resolution TEM analysis (HRTEM, not shown here), can be identified as $ZnMnO_3$ impurities. Consistently with the XRD data, it is found that when increasing the firing temperature they grow both in size and number. We should mention that the presence of small particles with a spinel $Zn_xMn_{3-x}O_4$-like structure (which is indeed close to that of $ZnMnO_3$) can not be discarded.

The microstructure observed in the case of Co-doped samples was similar to that shown in Figure 2. HRTEM analysis (not shown here) shows the presence of small grains (~10nm) of the spinel $Zn_xCo_{3-x}O_4$, and, more importantly, it was not observed any trace of metallic Co segregates.

Figures 3(a) and (b) show the magnetization as a function of the applied field corresponding to Co- and Mn-doped samples, respectively. Measurements were performed at room temperature, and the diamagnetic signal arising from the sample holder was carefully discounted in all cases. It is found that those samples fired at 300ºC (ZC1_a and ZM1_a) display a ferromagnetic behavior, while those treated at 600ºC (ZC1_d and ZM1_d) are fully paramagnetic. The extracted saturation magnetizations of ZC1_a and ZM1_a samples resulted $M_S \sim 0.01\mu_B$/Co and $\sim 0.007\mu_B$/Mn, respectively, indicating that only a small fraction of the transition metals present in the samples are ferromagnetically coupled, while the rest remain paramagnetic. In the first place, we should mention that the observed ferromagnetism does not seems to be related with the presence of $Zn_xCo_{3-x}O_4$ and $ZnMoO_3$ impurities, as it is evident that the samples fired at 600ºC, where the amount of these segregated phases is clearly higher, do not show any ferromagnetic coupling. Additionally, the extrinsic mechanisms proposed in Refs. [5,8] to explain the ferromagnetic behavior of Mn-doped ZnO can be reasonably discarded, as we have not observed the presence of any trace of $(ZnMn)_2O_3$ or $MnO_2$-like phases. In the case of Co-doped samples, our detailed XRD and TEM measurements also failed to detect the presence of any metallic Co aggregates. In consequence, it is reasonable to propose that the ferromagnetic coupling observed in our Co- and Mn-doped samples fired at low temperatures presents an intrinsic origin.

Interestingly enough, we have found that the ferromagnetic coupling is modified with time. Figure 4 shows two room temperature hysteresis loops performed on sample ZC1_a (stored at room temperature, in air) with a difference of about two months, being evident the presence of a "magnetic aging" mechanism: the initial ferromagnetic signal vanishes with time, leading to a paramagnetic behavior. It is worth noting that this change was not accompanied by any measurable structural or chemical modification, as the XRD patterns corresponding to the "fresh" and "aged" samples were completely identical. A similar behavior was found in the case of sample ZM1_a.

Although the microscopic origin of the observed phenomenology is not fully understood, it is reasonable to argue that the observed changes in the ferromagnetic interaction could be related to modifications on the point-defects structure of our samples, which can plausibly take place at room temperature along a period of some weeks. At this point, it is natural to ask if this behavior can be reversed, for example, by treating the samples at low temperatures in different atmospheres. Further experiments are in progress with enlightening results, and will be reported in a near future.

## 4. Conclusions

In summary, we have reported on the study of Co- and Mn-doped ZnO nanopowders prepared by a soft chemistry route. It was found that those samples fired at lower temperatures display a ferromagnetic coupling that was proposed as being intrinsic. The possible key role of point-defects on the stabilization of the ferromagnetic coupling of these kind of diluted magnetic semiconductors was signaled.

## 5. Acknowledgements

We acknowledge financial support from project MAT2005-5656-C04-01. Quality Chemicals would like to thank CIDEM-Generalitat de Catalunya for financial support RDITCRIND04-0153.

**Figure Captions**

**Figure 1**: XRD patterns corresponding to Mn- and Co doped ZnO, treated at temperatures of 300ºC and 600ºC. Diffraction peaks marked as * and o signal the presence of $Zn_xCo_{3-x}O_4$ and $ZnMoO_3$ impurities, respectively. The intensities are shown in a logarithmic scale.

**Figure 2:** (a) TEM image corresponding to the sample ZM1_a, showing the typical nanostructure of the studied powders; (b) TEM image of the same sample at a closer scale. The squared particle corresponds to a $ZnMnO_3$ aggregate.

**Figure 3:** Room temperature magnetizations as a function of the temperature of $Co_{0.1}Zn_{0.9}O$ (a) and $Mn_{0.1}Zn_{0.9}O$ (b) samples. The diamagnetic contribution of the sample holder was subtracted from the raw data.

**Figure 4:** Room temperature hysteresis loop corresponding to ZC1_a as-grown and "aged" sample. Measurements were performed with a ~2 months difference.

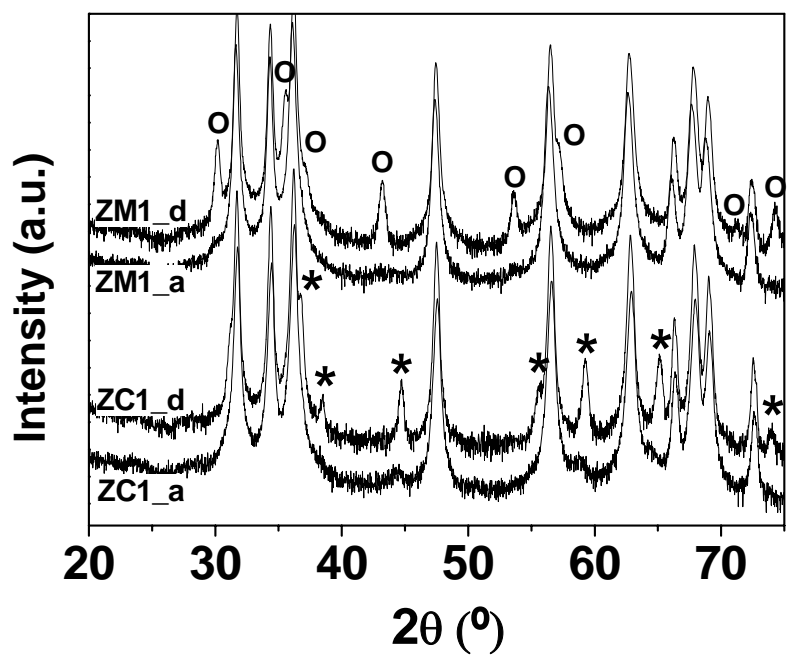

**Figure 1**

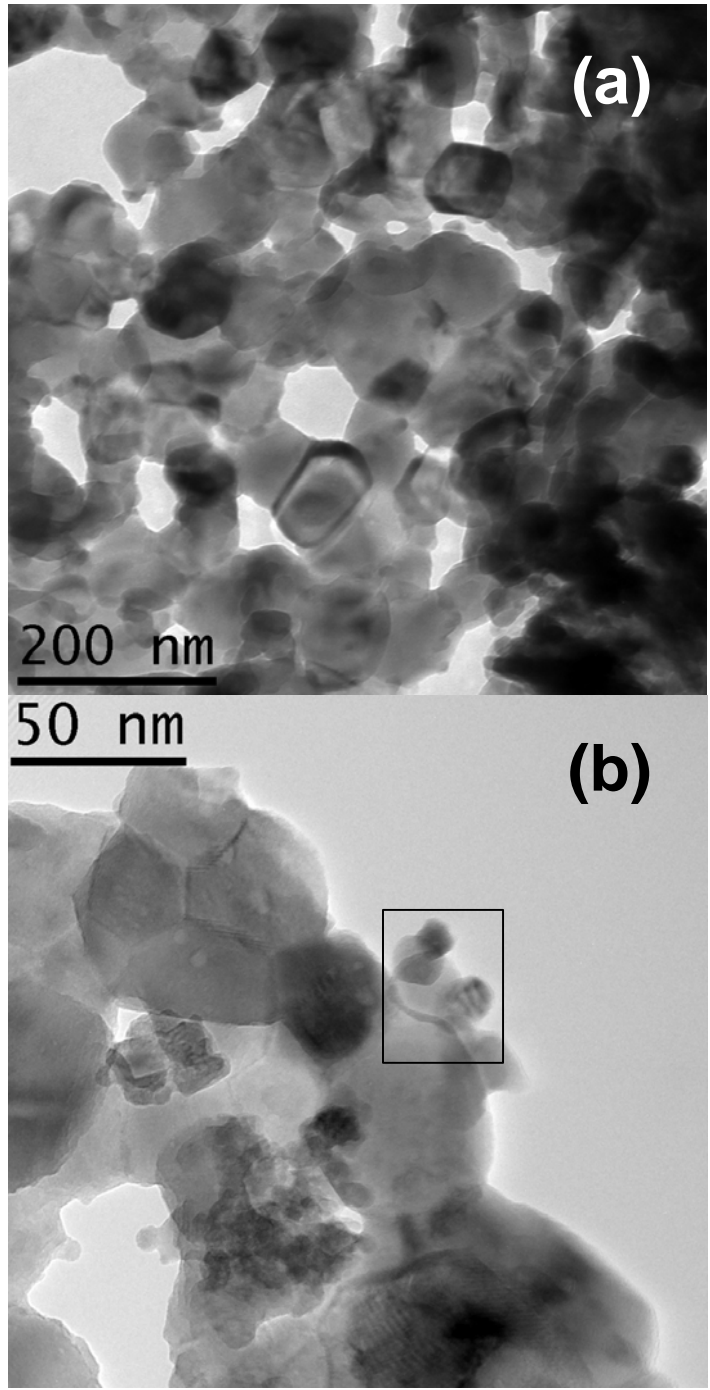

**Figure 2**

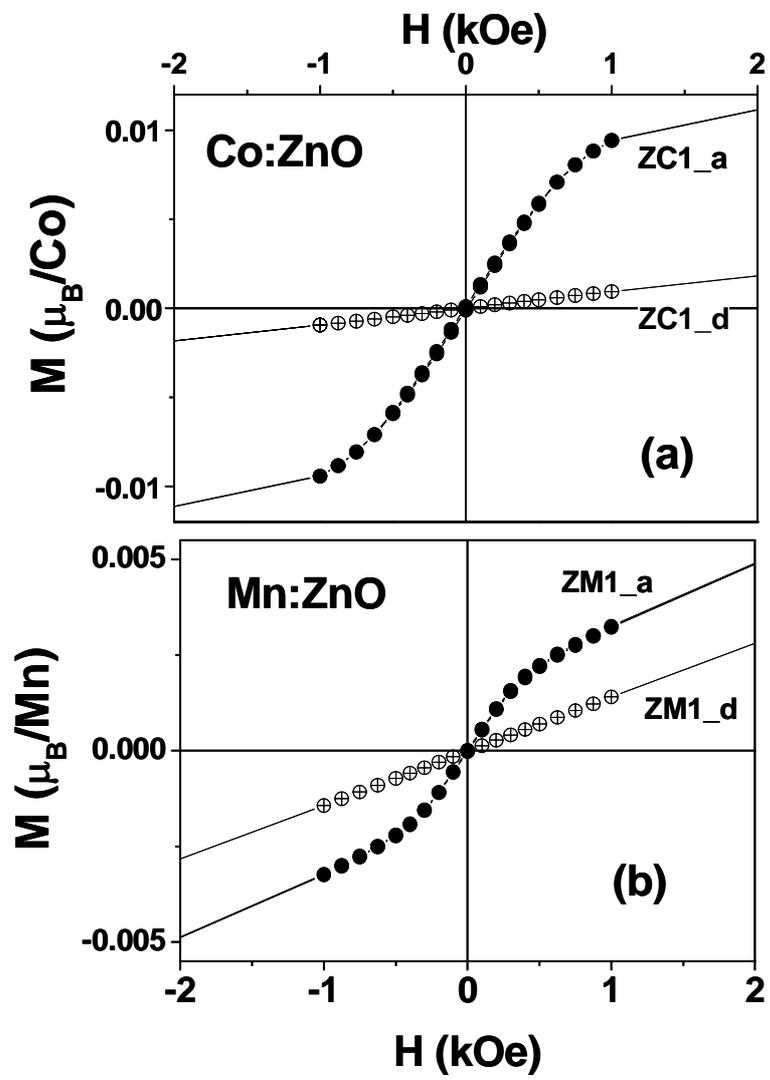

**Figure 3**

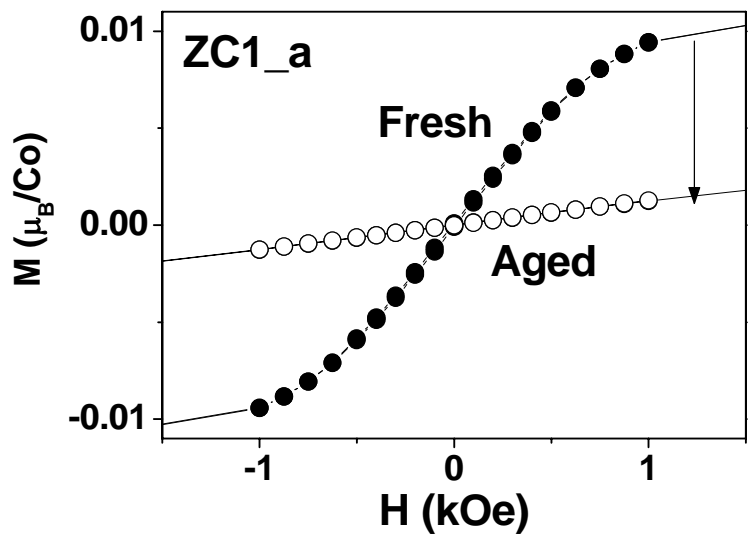

**Figure 4**